\documentclass[aps, prl, twocolumn, superscriptaddress]{revtex4}
\usepackage{amsthm, amssymb, verbatim,color}

\def\t{^{\mbox{\tiny T}}}
\def\bfe{{\bf e}}
\def\bfa{{\bf a}}
\def\bfb{{\bf b}}
\def\bfs{{\bf s}}

\def\a{{\cal A}}
\def\b{{\cal B}}
\def\e{{\rm e}}
\def\I{{\rm i}}
\def\mfh{\mathfrak{h}}
\def\g{{\cal G}}
\def\h{{\cal H}}

\def\reals{\mathbb{R}}
\newcommand{\eq}[1]{(\ref{#1})}
\def\vn{v}

\newcommand{\ket}[1]{|#1\rangle}

\newcommand{\proj}[1]{|#1\rangle\langle#1|}

\begin{document}

\title{Deriving quantum theory from its local structure and reversibility}

\author{Gonzalo de la Torre}
\affiliation{ICFO-Institut de Ci\`encies Fot\`oniques, E-08860 Castelldefels, Barcelona, Spain}
\author{Llu\'{\i}s Masanes}
\affiliation{ICFO-Institut de Ci\`encies Fot\`oniques, E-08860 Castelldefels, Barcelona, Spain}
\author{Anthony J. Short}
\affiliation{DAMTP, University of Cambridge, CB3 0WA, Cambridge, UK}
\author{Markus P. M\"uller}
\affiliation{Perimeter Institute for Theoretical Physics, 31 Caroline Street North, Waterloo, ON N2L 2Y5, Canada}

\date{\today}

\begin{abstract}
 We investigate the class of physical theories with the same local structure as quantum theory, but a potentially different global structure. It has previously been shown that any bipartite correlations generated by such a theory must be simulatable in quantum theory, but that this does not hold for tripartite correlations.  Here we explore whether imposing an additional constraint on this space of theories - that of dynamical reversibility - will allow us to recover the global quantum structure. In the particular case in which the local systems are identical qubits, we show that any theory admitting at least one continuous reversible interaction must be identical to quantum theory.
\end{abstract}

\maketitle

To gain a better understanding of quantum theory, it is helpful to compare and contrast it with other conceivable physical theories within a more general  framework \cite{hardy, gpt_barrett, MM}. In this paper, we will be  interested in those theories which have the same local structure as quantum theory, but a potentially different global structure. What singles out quantum theory from within this class of \emph{locally quantum} theories?

In a recent paper, Barnum et al \cite{barnum} showed that the set of bipartite correlations attainable in any
locally quantum theory is precisely the set of quantum correlations. However, such theories may in general contain non-quantum states,
corresponding to entanglement witnesses \cite{witnesses}. Furthermore, when three or more parties are considered, locally quantum theories can yield stronger than
quantum non-local correlations \cite{acinicfo}.

An important property of quantum theory which is not shared by all theories is: the existence of a continuous reversible transformation between any two pure states.
Demanding that the fundamental dynamics is reversible and continuous in time seems like a natural physical requirement, and is a key component of several recent axiomatic reconstructions of quantum theory \cite{hardy, chiribella, MM, DB}. Furthermore, an alternative theory yielding much stronger non-local correlations than quantum theory, known as box-world, was recently shown to contain no reversible interactions \cite{gross}. Here we explore how reversibility constrains the global structure of quantum theory.

Our main result is to show that, when the individual systems are identical qubits, the existence of any continuous reversible interaction is enough to single out quantum theory from all
possible locally quantum theories.

\bigskip\emph{Locally quantum theories.} In the general probabilistic framework of \cite{hardy, gpt_barrett, MM}, the state of a system is characterised by the outcome probabilities it yields for some set of  measurements, called \emph{fiducial} measurements. This set is generally non-unique, but must be sufficient to derive the outcome probabilties for any other measurement. For qubits, we will take the fiducial measurements to be the three Pauli spin operators, which we denote by $\sigma_1, \sigma_2, \sigma_3$.

Joint systems are assumed to be completely characterised by the probability distributions for every combination of fiducial measurements on the component systems.  This property is called \emph{local tomography}. The  state of $n$ qubits can therefore be represented by the conditional probability distribution $P(a_1, \ldots, a_n | x_1, \ldots, x_n)$
which gives the joint probability of obtaining results $a_1, \ldots, a_n \in \{+1,-1\}$ when measuring $x_1, \ldots, x_n \in \{\sigma_1, \sigma_2, \sigma_3 \}$ respectively on the $n$ systems. In order for the probabilities of outcomes on a single system to be well-defined, and to prevent instantaneous signalling between parties, it is important that
\begin{equation}\label{eq:nosig}
\sum_{a_k} P(a_1, \ldots, a_k, \ldots, a_n | x_1, \ldots, x_k, \ldots, x_n)
\end{equation}
is independent of $x_k$ for all $k$.

It is shown in \cite{acinicfo} that any  state satisfying (\ref{eq:nosig}) can be represented in an analogous way to a standard quantum state, as a trace-1 Hermitian operator $\rho$ on a $2^n$-dimensional Hilbert space, with the outcome probabilities for the fiducial measurements obtained in precisely the same way as in standard quantum theory. However, the operator $\rho$ need not be positive in general. As we are considering local quantum theories, the set of separable states and the set of local measurements must be the same as in quantum theory. However, the entangled states and measurements may differ.

Due to their action on probabilistic mixtures of states, transformations must be represented as linear maps on $\rho$~\cite{hardy, gpt_barrett}. Note that we adopt an operational approach, so the allowed transformations include possibilities in which an ancilla is prepared, evolves jointly with the system, and is then discarded (in  quantum theory  the allowed transformations are the completely-positive trace-preserving (CPTP) maps). In what follows we will be particularly interested in those transformations which are reversible and connected to the identity, such that they could be implemented continuously in time. The connected reversible transformations on $n$ qubits form a group $\g$.

The group $\g$ must obey two important conditions. The first is that it should contain the local unitary transformations $\g_{\rm loc}$ for the qubits. A unitary transformation $U \in {\rm SU}(d)$ acts on a $d$-level quantum system via the adjoint action ${\rm ad}_U [\rho] = U\rho U^\dagger$. Hence for $n$ qubits
\[
	\g_{\rm loc} = \{
		{\rm ad}_{U_1} \otimes \cdots \otimes {\rm ad}_{U_n} : U_r \in {\rm SU} (2)
	\} \subseteq \g.
\]
The second condition on the group $\g$ is that the combination of preparing a product state, transforming it using any $G \in \g$, then performing a product measurement must yield a valid outcome probability. These two conditions allow us to prove our main technical result, the proof of which will be given later in the paper.

\bigskip {\bf Theorem 1.}\textit{
Let $\g$ be a connected group which acts linearly on the set of $2^n \times 2^n$ Hermitian matrices and satisfies:}
\begin{enumerate}
	\item $\g_{\rm loc} \subseteq \g$,
	
	\item ${\rm tr} \Big( (\mu_1 \otimes \cdots \otimes \mu_n)\,
		G[\rho_1 \otimes \cdots \otimes \rho_n] \Big) \in [0,1]$
		\textit{for any $G\in\g$ and any qubit states $\mu_r$ and $\rho_r$.}
\end{enumerate}
\textit{If $\g \neq \g_{\rm loc}$ then there exist: $G\in\g$, a re-ordering of the qubits, an entangling unitary $U\in {\rm SU}(4)$, and a single-qubit state $\sigma$, such that one of the following possibilities holds:}
\begin{enumerate}
	\item $G[\rho_{12} \otimes \sigma^{\otimes (n-2)}] = {\rm ad}_U [\rho_{12}] \otimes \sigma^{\otimes (n-2)}$,

	\item $G[\rho_{12} \otimes \sigma^{\otimes (n-2)}] = (T_1 \circ {\rm ad}_U \circ T_1) [\rho_{12}] \otimes \sigma^{\otimes (n-2)}$,
\end{enumerate}
\textit{for any Hermitian matrix $\rho_{12}$, where $T_1$ is the partial transposition operation on qubit 1.}

\bigskip This theorem means that in any locally quantum theory of $n$ qubits which allows for (not necessarily quantum) interaction, $\g \neq \g_{\rm loc}$, there exist a particular pair of qubits, $i$ and $j$, on which we can implement either the quantum transformation $G^{U}_{ij}={\rm ad}_U$, or the non-quantum $H^{U}_{ij}= (T_i \circ {\rm ad}_U \circ T_i) $. As any bipartite entangling unitary plus local unitaries are sufficient to generate all unitary transformations \cite{aram}, when we can implement $G^U_{ij}$ we can implement all unitary transformations on qubits $i$ and $j$. If we can  implement $H^U_{ij}$ instead, we note that for any $V_i \in SU(2)$, there exists $V_i' \in SU(2)$ such that ${\rm ad}_{V'_i} = (T_i  \circ {\rm ad}_{V_i} \circ T_i)$. Hence by sequences of local transformations and $H^U_{ij}$ operations we can implement $(T_i \circ {\rm ad}_V \circ T_i)$ on qubits $i$ and $j$, for any $V \in SU(4)$ .

Now consider the additional assumption that all qubits are identical, in the sense that they are the same type of system (although they may have different states). This means that given $(n+m)$ qubits (for any $m\geq 0$), for any $G\in \g$, and any permutation $\pi$ of the qubits, $\pi \circ (G \otimes I^{\otimes n}) \circ \pi^{-1} $ is an allowed transformation. In this case we can prove a stronger result.

\bigskip {\bf Theorem 2.}\emph{ Consider any locally-tomographic theory in which the individual systems are identical qubits. If the theory admits any continuous reversible interaction between systems, then the allowed states, measurements, and transformations must be identical to those in quantum theory.}

\bigskip \emph{Proof.} The existence of at least one continuous reversible interaction implies that we can find  $n \geq2$ qubits such that  $\g \neq \g_{\rm loc}$ and Theorem 1 holds. Furthermore, as the systems are identical we can perform either $G^U$ or $H^U$ between any pair of qubits.

We now show that the ability to implement one (and thus all) $H^U$ between any two qubits is inconsistent. Note that by acting on three qubits with a sequence of local unitaries, $H^U_{12}$ and $H^U_{13}$ transformations we could implement $(T_1 \circ {\rm ad}_V \circ T_1) $ for any  $V \in SU(8)$. However, this includes the transformations $G^U_{23}$.  If this were possible, we could first prepare a state $\rho_{23}$ with a negative eigenvalue using $H^U_{23}$ and $\g_{\rm loc}$  (e.g. by implementing $(T_2 \circ {\rm ad}_V \circ T_2) [\proj{00}] $  when $V\ket{00} = \frac{1}{\sqrt{2}} [\ket{00} + \ket{11}]$), then map the negative eigenvector onto the $\ket{00}$ state using $G^U_{23}$ and $\g_{\rm loc}$. The final state would assign a negative value to the probability of obtaining $00$ in a local computational basis measurement, which is inconsistent.

The only remaining possibility is that we can implement $G^U$ between any two qubits. In conjunction with local transformations, this allows us to implement any unitary transformation. Given that we can also perform local preparations and measurements, this allows us to create any quantum state, and to implement any CPTP map or quantum measurement. We now rule out any other states, measurements, and transformations. Note that any state $\rho$ with a negative eigenvalue would assign a negative probabilty for the outcome of some quantum measurement, and is therefore inconsistent. Non-quantum measurement effects can be ruled out similarly. Finally, note that transformations must be completely positive maps, or they could be used to generate a state with a negative eigenvalue by acting on part of some entangled state. As all transformations must be trace-preserving, they must be CPTP maps. Hence the set of allowed states, measurements and transformations is precisely that of standard quantum theory. $\Box$

\bigskip 
The above proof is similar to an argument from~\cite{DB}. To help us prove Theorem 1, we first discuss an alternative representation of states based on Bloch-vectors, and the Lie algebra of  $\g$.

\bigskip\emph{A multi-qubit Bloch-vector representation.} As we are considering qubits, it is helpful to adopt a generalised Bloch vector representation of the state. We expand $\rho$ as
\begin{equation}\label{rho Pauli}
	\rho =\  2^{-n}\!\!\!\! \sum_{\alpha_1, \ldots, \alpha_n} \!
	r_{\alpha_1 \cdots \alpha_n} \
	\sigma_{\alpha_1} \otimes \cdots \otimes \sigma_{\alpha_n},
\end{equation}
where $\alpha_k \in \{0,1,2,3\}$ and $\sigma_0$ is the identity operator. For clarity, we will use the convention throughout that $\alpha, \beta, \gamma \in \{0,1,2,3\}$ and $i,j \in \{1,2,3\}$.

The real vector
\begin{equation}\label{rho real}
	r_{\alpha_1 \cdots \alpha_n} =  {\rm tr} \Big(
	(\sigma_{\alpha_1} \otimes \cdots \otimes \sigma_{\alpha_n})\, \rho
	\Big).
\end{equation}
is a complete representation of the state, and is related to the probability distribution $P(a_1, \ldots, a_n | x_1, \ldots, x_n)$
by an invertible linear map.

For a single qubit $r_0 =1$ and $r_i$ is the Bloch vector. Similarly, $n$-qubit product states can be represented by $r$-vectors of the form
\begin{equation}
	{r}=
	\vn(\bfa_1, \ldots , \bfa_n) =
	\left[\begin{array}{c} 1 \\ \bfa_1 \end{array} \right]
	\otimes \cdots \otimes
	\left[\begin{array}{c} 1 \\ \bfa_n \end{array} \right] .
\end{equation}
where the $\bfa_k$ are Bloch vectors ($\bfa_k \in \reals^3$ and $|\bfa_k| \leq 1$). As we are considering a locally quantum theory,
all these states are elements of the global state space.

Similarly, each measurement outcome can be associated with an \emph{effect} vector $p$, such that  the probability of getting that outcome when measuring the state $r$ is  $p\t r$.  The vectors $p=2^{-n}\vn(\bfb_1, \ldots , \bfb_n)$, where the  $\bfb_k$ are  Bloch vectors, all correspond to allowed product effects.

Transformations are represented by matrices acting on the state vector. In particular, the transformation $ \rho \rightarrow G[\rho]$ is represented by the matrix
\[
	H^{\alpha_1 \cdots \alpha_n}_{\beta_1 \cdots \beta_n} =
	\frac{1}{2^n} {\rm tr}
	\Big( (\sigma_{\beta_1} \otimes \cdots \otimes \sigma_{\beta_n})\,
	G[\sigma_{\alpha_1} \otimes \cdots \otimes \sigma_{\alpha_n}] \Big).
\]
which acts on the $r$ vector as $r \rightarrow H r$. The single qubit unitaries form a group with a simple matrix representation
\begin{equation}\label{h1}
	\h_q = \left\{
	\left[ \begin{array}{c|ccc}
		1 & 0 & 0 & 0\\
		\hline
		0 &&&\\
		0 & & R & \\
		0 &&&
	\end{array} \right]: R \in {\rm SO}(3) \right\}.
\end{equation}

We will denote the analogues of $\g$ and $\g_{\rm loc}$ in this representation by $\h$ and $\h_{\rm loc}$ respectively.

Since $\h$ is connected, there is a Lie algebra $\mfh$ such that, for each $H \in\h$ there is a matrix $X\in \mfh$ satisfying $H = \e^X$.
The standard quantum Lie algebra is the real vector space of traceless anti-Hermitian matrices, with basis
\[
   \{\I (\sigma_{\gamma_1} \otimes \ldots \otimes \sigma_{\gamma_n}) \,\,|\,\,
   (\gamma_1,\ldots,\gamma_n)\neq (0,\ldots, 0)\}.
\]
which act on the state through the commutator $\rho \to [\I (\sigma_{\gamma_1} \otimes \cdots \otimes \sigma_{\gamma_n}), \rho]$. When $n=2$, the matrix representation
of any basis element $X\in\mfh$ is
\[
		X^{\alpha_1 \alpha_2}_{\beta_1 \beta_2}
	=
		\frac{1}{2^2} {\rm tr}\Big(
		(\sigma_{\beta_1} \otimes \sigma_{\beta_2})
		\Big[
			\I (\sigma_{\gamma_1} \otimes \sigma_{\gamma_2}) ,
			(\sigma_{\alpha_1} \otimes \sigma_{\alpha_2})
		\Big]
	\Big) .
\]

Defining $I$ as the $4 \times 4$ identity matrix, and
\begin{equation}\label{in a}
\begin{array}{cc}
	A_\bfa =
	\left[
	\begin{array}{c|ccc}
		0 & 0 & 0 & 0\\
		\hline
		0 & 0 & a_3 & -a_2 \\
		0 & -a_3 & 0 & a_1 \\
		0 & a_2 & -a_1 & 0
	\end{array}
	\right],
&
	B_\bfa =
	\left[
	\begin{array}{c|ccc}
		0 & a_1 & a_2 & a_3\\
		\hline
		a_1 & 0 & 0 & 0 \\
		a_2 & 0 & 0 & 0 \\
		a_3 & 0 & 0 & 0
	\end{array}
	\right],

\end{array}
\end{equation}
we can write the matrix representation of the Lie algebra element $\I (\sigma_i \otimes \sigma_j)$ as
\begin{equation}\label{W_ij}
	X = 2 A_{\bfe_i} \otimes B_{\bfe_j} + 2 B_{\bfe_i} \otimes A_{\bfe_j},
\end{equation}
where $\bfe_i$ is the unit vector in the $i$-direction. The element $\I (\sigma_i \otimes \sigma_0)$ has the matrix representation
$ X= 2 A_{\bfe_i} \otimes I.$

\bigskip\emph{Proof of Theorem 1.} In the Bloch-representation described above, condition 2 from Theorem 1 is
\begin{equation}\label{c1}
	2^{-n} \vn (\bfb_1 ,\ldots , \bfb_n)\t H \vn (\bfa_1, \ldots , \bfa_n) \in [0,1],
\end{equation}
for all Bloch vectors $\bfa_r, \bfb_r$. Considering a group element close to the identity, $H=\e^{\epsilon X} \in \h$, and expanding equation \eq{c1} to second order in $\epsilon$ gives
\begin{equation} \label{expansion}
\vn (\bfb_1  \ldots  \bfb_n)\t \!\!\left(
	I^{\otimes n} + \epsilon X +\frac{\epsilon^2}{2} X^2
	\right)\! \vn (\bfa_1 \ldots  \bfa_n)\in [0,2^n]
\end{equation}
When all the Bloch vectors have unit length, we can use this expansion to derive the first-order constraints
\begin{equation}
\label{1oc}
\mathcal{C}[\bfa_1] \equiv	\vn(-\bfa_1, \bfb_2, \ldots , \bfb_n)\t X
	\vn(\bfa_1 , \bfa_2,\ldots , \bfa_n) =0,
\end{equation}
 and the second-order constraints
\begin{eqnarray}\label{2oc1}
	\vn(-\bfa_1, \bfb_2, \ldots , \bfb_n)\t X^2
	\vn(\bfa_1 , \bfa_2,\ldots , \bfa_n) &\geq& 0,
\\\label{2oc2}
	\vn(\bfa_1, \bfa_2, \ldots , \bfa_n)\t X^2
	\vn(\bfa_1 , \bfa_2,\ldots , \bfa_n) &\leq& 0.	
\end{eqnarray}
These constraints hold for all  $X\in \mfh$.

We initially use the first-order constraints in (\ref{1oc}). Considering $\mathcal{C}[\bfe_i] \pm \mathcal{C}[ -\bfe_i]=0$, we find
\begin{equation}\label{1oc1}
\begin{array}{ccc}

	X^{i \alpha_2 \cdots \alpha_n}_{i \beta_2 \cdots \beta_n}
	= 	
	X^{0 \alpha_2 \cdots \alpha_n}_{0 \beta_2 \cdots \beta_n}
& \textrm{and} &
	X^{i \alpha_2 \cdots \alpha_n}_{0 \beta_2 \cdots \beta_n}
	= X^{0 \alpha_2 \cdots \alpha_n}_{i \beta_2 \cdots \beta_n} .
\end{array}
\end{equation}
for all values of $i, \alpha_r, \beta_r$, where we have used the fact that the vectors $\vn(\bfa_2,\cdots , \bfa_n)$ linearly span the whole of $(\reals^4)^{\otimes (n-1)}$.

Similarly,  taking $\mathcal{C}[\frac{1}{\sqrt{2}} (\bfe_i + \bfe_j)] + \mathcal{C}[-\frac{1}{\sqrt{2}} (\bfe_i + \bfe_j)] =0$  for $i \neq j$, and using  (\ref{1oc1}), we obtain
\begin{equation} \label{1oc2}
X^{i \alpha_2 \cdots \alpha_n}_{j \beta_2 \cdots \beta_n}
	= - X^{j \alpha_2 \cdots \alpha_n}_{i \beta_2 \cdots \beta_n}.
\end{equation}

Let $\a$ and $\b$ denote the linear spans of the $A_\bfa$ and $B_\bfa$ matrices defined in \eq{in a} respectively, and let ${\cal I}$ denote the linear span of $I$. Then equations (\ref{1oc1}-\ref{1oc2}), together with the equivalent equations for the other $(n-1)$ qubits,
imply that $\mfh$ is a subspace of $(\a \oplus \b \oplus {\cal I})^{\otimes n}$. We equip $(\a \oplus \b \oplus {\cal I})$ with the standard matrix inner-product $\langle A,B \rangle={\rm tr}(A\t B)$, under which $\{A_{\bfe_1}, A_{\bfe_2}, A_{\bfe_3}, B_{\bfe_1},  B_{\bfe_2}, B_{\bfe_3}, I\}$ form an orthogonal basis for the space.

It is easy to see that the local Lie algebra is
\begin{equation}\label{h loc}
	\mfh_{\rm loc} =
	\bigoplus_\pi \pi (\a \otimes {\cal I}^{\otimes (n-1)}) ,
\end{equation}
where $\pi$ runs over all permutations of the $n$ factor spaces. Condition 1 from  Theorem 1 implies $\mfh_{\rm loc} \subseteq \mfh$.

If $X \notin \mfh_{\rm loc}$ then we can always re-order the subsystems such that $X$ has support on
the subspace $\mathcal{S} = \mbox{$\a^{\otimes n_A} \otimes \b^{\otimes n_B} \otimes {\cal I}^{\otimes n_I}$}$,
where $n_A +n_B +n_I = n$, and at least one of the inequalities, $n_A\geq 2$ or $n_B \geq 1$, holds.
In particular, there is a matrix $M_S=A_{\bfa_1} \otimes\ldots\otimes A_{\bfa_{n_A}}\otimes B_{\bfb_1}\otimes\ldots\otimes B_{\bfb_{n_B}} \otimes I^{\otimes n_I}\in \mathcal{S}$
that has non-zero overlap with $X$.
Furthermore, there exists a local transformation $H_{\rm loc} \in \h$ such that $H_{\rm loc} M_S H_{\rm loc}^{-1} \propto A^{\otimes n_A}_{\bfe_1} \otimes B^{\otimes n_B}_{\bfe_1} \otimes I^{\otimes n_I}$.

Let us define $E_0 = A_{\bfe_1}$ and $E_1 = B_{\bfe_1}$, and denote the linear span of these two matrices by  ${\cal E}$. Then $H_{\rm loc} X H_{\rm loc}^{-1}$ has support on ${\cal E}^{\otimes m} \otimes {\cal I}^{\otimes n_I}$, where $m=n_A +n_B$.

 For any matrix $M \in (\a \oplus \b \oplus {\cal I})$, the  projector onto ${\cal I}$ is given by  $\Phi_{\cal I} [M] = \int_{\h_q}\!\! dH\, H M  H^{-1}$, where $\h_q$ is the group of all single qubit unitaries defined in (\ref{h1}). Similarly, the  projector onto ${\cal E}$ is given by $\Phi_{\cal E} [M]  = \int_{\h_{\bfe_1}} \!\! dH\, H M  H^{-1}- \Phi_{\cal I}[M]$ where $\h_{\bfe_1} = \{H\in \h_q ; H \vn(\bfe_1) = \vn(\bfe_1)\}$. As $HXH^{-1} \in\mfh$ whenever $X\in \mfh$ and $H\in \h$, it follows that the matrix
 \begin{equation}
	Y =
	(\Phi_{\cal E}^{\otimes m} \otimes \Phi_{\cal I}^{\otimes n_I}) \big[ H_{\rm loc} X H_{\rm loc}^{-1} \big]
\end{equation}
is a non-zero element of $\mfh$, and fully contained in $\mbox{${\cal E}^{\otimes m} \otimes {\cal I}^{\otimes n_I}$}$. This allows us to expand $Y$ as
  \begin{equation}
	Y =	\sum_{\bfs \in \{0,1\}^m}\!\!\!\! c_{\bfs} \
	E_{s_1} \otimes \cdots \otimes E_{s_m} \otimes I^{\otimes n_I},
\end{equation}
Since $E_0 E_1 = E_1 E_0 =0$ we have
\begin{equation}\label{Xpp2}
	Y^2 =
	\sum_{\bfs \in \{0,1\}^{m}}\!\!\!\! c_{\bfs}^2 \
	E_{s_1}^2 \otimes \cdots \otimes E_{s_{m}}^2 \otimes I^{\otimes n_I}.
\end{equation}
Also note that
\begin{eqnarray}\label{eq20}
	&\vn(\pm \bfe_2)\t E_0^2\, \vn(\bfe_2) = \mp 1,
\quad
	&\vn(\pm \bfe_2)\t E_1^2\, \vn(\bfe_2) =1, \quad
\\ \label{eq21}
	&\vn(\bfe_1)\t E_0^2\, \vn(\bfe_1) =0,
\quad
	&\vn(\bfe_1)\t E_1^2\, \vn(\bfe_1) =2. \quad
\end{eqnarray}


We now use these equations, together with the second order constraints given by (\ref{2oc1}) and (\ref{2oc2}), to derive some properties of the coefficients $c_{\bfs}$.
First note that according to~(\ref{2oc2})
\begin{equation}
	\vn(\bfe_1, \ldots, \bfe_1)\t Y^2
	\vn(\bfe_1, \ldots, \bfe_1) = c_{(1,1,1,\ldots , 1)}^2\, 2^n \leq 0
\end{equation}
hence $c_{(1,1,1,\ldots , 1)}=0$. This also rules out the case $n_A=0, n_B=1$, which implies $m \geq 2$.

Now consider the two inequalities
\begin{eqnarray}
\label{i1}
	\vn (-\bfe_2, \bfe_2, \bfe_1, \ldots ,\bfe_1)\t  Y^2
	\vn ( \bfe_2, \bfe_2, \bfe_1, \ldots ,\bfe_1) \geq 0,
\\ \label{i2}
	\vn (\bfe_2, -\bfe_2, \bfe_1, \ldots ,\bfe_1)\t  Y^2
	\vn ( \bfe_2, \bfe_2, \bfe_1, \ldots ,\bfe_1) \geq 0,
\end{eqnarray}
derived from (\ref{2oc1}). Adding these two inequalities together we obtain $(c_{(1,1,1,\ldots , 1)}^2 -c_{(0,0,1,\ldots , 1)}^2) 2^{n-1} \geq 0$, which implies $c_{(0,0,1,\ldots , 1)}=0$.
After removing the terms which are zero in these two inequalities we obtain
\begin{eqnarray*}
	 ( c^2_{(0,1,1,\ldots, 1)} -c^2_{(1,0,1,\ldots, 1)} ) 2^{n-2}  \geq 0,
\\
	 ( c^2_{(1,0,1,\ldots, 1)} -c^2_{(0,1,1,\ldots, 1)} ) 2^{n-2}  \geq 0,
\end{eqnarray*}
respectively. Together these imply $c_{(1,0,1,\ldots, 1)} = \pm c_{(0,1,\ldots, 1)}$.

We now show by induction that  $ c_{(0,1, \ldots, 1)} \neq 0$.  From the construction of $Y$, it is clear that $c_{(0 \ldots 0, 1, \ldots 1)} \neq 0$ when the index contains $n_A$ zeroes. Now, for some $l \geq 2$, suppose that
$c_{(s_1,\ldots, s_{l},1,\ldots, 1)} \neq 0$ if and only if  $s_1 = \dots = s_l =0$.  In this case, the constraint
\[
\vn (\pm\bfe_2, - \bfe_2 \ldots -\bfe_2, \bfe_1 \ldots\bfe_1)\t Y^2 \vn (\bfe_2 \ldots \bfe_2, \bfe_1 \ldots \bfe_1) \geq 0 \label{eq19}	
\]
implies $\mp c_{(0 \ldots 0,1 \ldots 1)}^2 2^{n-l} \geq 0$, which is impossible. Hence, there must exist another component  $c_{(s_1,\ldots, s_{l},1,\ldots, 1)} \neq 0$ for which the index contains less than $l$ zeroes.  Rearranging the first $l$ qubits such that the zeroes occur at the start of $\bfs$ and proceeding by induction, we find that either $c_{(0,1,\ldots , 1)} \neq 0$ or $c_{(1,1,\ldots , 1)} \neq 0$. However, the latter possibility is ruled out above.

Taking the $n-2$ leftmost qubits to be in the state $\sigma = \vn(\bfe_1)$, and noting that $E_0 \sigma = 0$ and $E_1 \sigma = \sigma$, the element $Z = c_{(0,1,1,\ldots, 1)}^{-1} Y \in \mfh$ acts as
\[
	Z \Big( {\bf r}_{12} \otimes \sigma^{\otimes (n-2)} \Big) = \big((A_{\bfe_1} \otimes B_{\bfe_1} \pm B_{\bfe_1} \otimes A_{\bfe_1}) {\bf r}_{12}\big)
\otimes
	\sigma^{\otimes (n-2)}
\]
for any vector ${\bf r}_{12}$. In the $+$ case, the action on the first two qubits is that of the quantum Lie algebra element $i(\sigma_1 \otimes \sigma_1)$, which can be used to generate the entangling  unitary transformation $U = e^{i (\sigma_1 \otimes \sigma_1)}$. In the $-$ case,  the Lie algebra element on the first two qubits can be written as $-T_1 (A_{\bfe_1} \otimes B_{\bfe_1} + B_{\bfe_1} \otimes A_{\bfe_1}) T_1$, where $T_1$ is the transpose operation on the first qubit. This can be used to generate the transformation $T_1 \circ {\rm ad}_U \circ T_1$. Rewriting these results in terms of $G[\rho]$, we recover Theorem 1. $\Box$

\bigskip\emph{Conclusions.} We have shown that quantum theory is the only theory in which (i) local systems behave like identical qubits (ii) there exists at least one continuous, reversible interaction. This highlights the importance of dynamical reversibility to the non-local structure of quantum theory.

Because we use Bloch-spheres and Lie algebras in our proof, it does not easily generalise to higher dimensional systems or discrete transformations. However, we conjecture that quantum theory is the only theory which is locally quantum, and in which there exists a reversible interaction between any pair of systems. More generally, an interesting open question is whether all reversible, locally-tomographic theories can be represented within quantum theory.

\bigskip\emph{Acknowledgments.} The authors thank Jonathan Oppenheim for helpful discussions. LM acknowledges support from CatalunyaCaixa. AJS acknowledges support from the Royal Society. GdlT acknowledges support from Spanish FPI grant (FIS2010-14830). Research at Perimeter Institute is supported by the Government of Canada through Industry Canada and by the Province of Ontario through the Ministry of Research and Innovation.


\begin{thebibliography}{99}

\bibitem{hardy} L.Hardy, {\em Quantum theory from five reasonable axioms}, arXiv:quant-ph/0101012.


\bibitem{gpt_barrett}  J. Barrett, Phys. Rev. A, \textbf{75}, 032304 (2007)

\bibitem{MM} Ll. Masanes and M. M\"uller, New J. Phys. \textbf{13}, 063001 (2011).

\bibitem{barnum} H. Barnum, S. Beigi, S. Boixo, M. B. Elliott, and S. Wehner, Phys. Rev. Lett. \textbf{ 104}, 140401 (2010).

\bibitem{witnesses} M. Horodecki, P. Horodecki, and R. Horodecki, Phys. Lett. A \textbf{223}, 1 (1996).

\bibitem{acinicfo} A. Ac\'{\i}n, R. Augusiak, D. Cavalcanti, C. Hadley, J. K. Korbicz, M. Lewenstein, Ll. Masanes, and M. Piani ,  Phys. Rev. Lett. \textbf{104}, 140404 (2010)

\bibitem{DB} B.\ Daki\'c, C.\ Brukner, {\em Quantum Theory and Beyond: Is Entanglement Special?}, arXiv:0911.0695v1.

\bibitem{chiribella} G. Chiribella, G. M. D'Ariano and P. Perinotti,  Phys. Rev. A \textbf{84}, 012311 (2011).

\bibitem{gross} D. Gross, M. M\"uller, R. Colbeck and O. C. O. Dahlsten,  Phys. Rev. Lett. \textbf{104}, 080402 (2010).

\bibitem{aram} A. Harrow, Q. Inf. Comp. \textbf{8}, 715 (2008).

\end{thebibliography}
\end{document}